\documentclass[pra,a4paper,twocolumn,superscriptaddress,showpacs,showkeys]{revtex4}

\usepackage{latexsym}
\usepackage{dcolumn}
\usepackage{bm}
\usepackage{upgreek}
\usepackage{amsmath}
\usepackage[latin1]{inputenc}
\usepackage[normalem]{ulem}

\usepackage{color}
\usepackage{subfigure}
\usepackage{paralist}
\usepackage{xcolor}

\usepackage{graphicx}
\graphicspath{{./figs_rev/}}

\newcommand {\queq}[1]{(\ref{#1})}
\newcommand {\qeq}[1]{Eq.~\queq{#1}}

\newcommand {\Rgrain}{R_{\rm grain}}
\newcommand {\Rred}{R_{\rm red}}
\newcommand {\xiyield}{\xi_{\rm yield}}

\newcommand {\beql}[1]{\begin{equation} \label{#1}}
\newcommand {\eeql}{\end{equation}}
\newcommand {\beq}{\begin{equation}}
\newcommand {\eeq}{\end{equation}}

\begin{document}

\date{\today}

\title{ Chondrule dust rim growth: Influence of restructuring using molecular dynamics simulations
 }
\author{Chuchu Xiang}
\affiliation{%
Center for Astrophysics, Space Physics and Engineering Research, Baylor University, Waco, TX 76798-7316, USA;
}

\author{Nina Merkert (n\'{e}e Gunkelmann)}
\affiliation{%
Clausthal University of Technology, Institute of Metallurgy, Arnold-Sommerfeld-Str.~6, 38678 Clausthal-Zellerfeld, Germany}

\author{Lorin S. Matthews}
\affiliation{%
Center for Astrophysics, Space Physics and Engineering Research, Baylor University, Waco, TX 76798-7316, USA;
}

\author{Augusto Carballido}
\affiliation{%
Center for Astrophysics, Space Physics and Engineering Research, Baylor University, Waco, TX 76798-7316, USA;
}

\author{Truell W. Hyde}
\affiliation{%
Center for Astrophysics, Space Physics and Engineering Research, Baylor University, Waco, TX 76798-7316, USA;
}

\begin{abstract}

We investigate the influence of disruptive collisions on chondrule rim growth, emphasizing the role of kinetic energy in determining the outcomes of these interactions. We establish a threshold of approximately 10 cm/s for the "hit-and-stick" collision regime, beyond which significant changes occur in the structure of rimmed chondrules. Our findings highlight that at low collision energies (KE $< 10^{-12}$ J), minimal structural alteration takes place, while higher energies (KE up to $10^{-10}$ J) lead to compaction of the rim, reducing both its thickness and porosity. Collisions with energies exceeding $10^{-8}$ J result in the complete disruption of the rim, with particles being expelled from it. These results are correlated with the turbulence levels within the disk, as kinetic energy scales with the relative velocities of colliding particles. Leveraging machine learning models trained on our collision data, we predict changes in rim characteristics and employ these predictions in a Monte Carlo simulation to explore rim growth dynamics. Our simulations reveal that rim development is sustained in low-turbulence environments ($\alpha \leq$ $10^{-5}$), while intermediate turbulence levels ($\alpha$ = $10^{-3}$ to $10^{-4}$) lead to erosion, preventing further rim accumulation in high-turbulence contexts. 

\end{abstract}



\maketitle


\section{Introduction}

Dust grains are known to be the fundamental material from which planets eventually form through agglomeration processes in protoplanetary disks composed of gas and dust \citep{Blu10}. Fragments of planetesimals, often retrieved as meteorites on Earth, are frequently observed to contain chondrules. These round glassy objects, which were once molten, became incorporated into meteorite parent bodies. \citet{Reipurth2007protostars} provide a comprehensive review of dust aggregation and include relevant findings concerning chondrules.

The exact process of chondrule formation remains not fully understood. It is believed that chondrules formed through rapid heating of dust aggregates followed by quenching. One hypothesis suggests that chondrules are created from cooling material ejected during collisions of large objects in protoplanetary disks, such as planetesimals and protoplanets. \citet{JMMZ15}, for example, simulated impacts and concluded that such events produce enough chondrules to account for observed abundances. Further simulations and investigations into the potential production of chondrules through collisions have been conducted by \citet{HWMO16} and \citet{WMOH17}. 

Once formed, chondrules can acquire a rim of dust grains. It is often assumed that the formation of these fine-grained rims (FGR) takes place in the gas of the solar nebula. Assuming the nebular hypothesis for rim formation is true, the structure of FGRs can reveal crucial details about the properties of the solar nebula during FGR formation, such as gas velocities, turbulent viscosities, and ionization levels, while also establishing the foundation for subsequent stages of planetesimal formation, including collisions between aggregates of rimmed chondrules and potential dust rim compaction by nebular shock waves.  \citet{OCT08} and \citet{XCH19} conducted Monte Carlo simulations of chondrules accumulating dust to form a rim. \citet{OCT08} found that this dust rim could facilitate the sticking together of chondrules in subsequent collisions. \citet{XCH19} determined that the porosity of the resulting rim decreases in more turbulent formation environments. Recently, the authors extended their results to the study of non-spherical (ellipsoidal) dust grains \cite{XCMH23}.
\citet{Bre93} examined the chemical composition and microstructure of matrix material and chondrule rims in the chondrite ALHA77307, finding results that support the nebular dust accretion model. \citet{HK18} used X-ray tomography on the Murchison meteorite to study the morphology of chondrules and the thicknesses of their dust rims, suggesting formation in a nebula. 
Fabric analysis, such as that by \citet{BHP11} on the chondrite matrix and chondrule rims, also supports a model where chondrules accreted a dust rim that was subsequently compressed, with the resulting bodies then accreted into a chondrite. 
\citet{BGN13} conducted experiments using an impact setup with a cylindrical aluminum projectile and a target material similar to chondrites. They determined the dynamic pressure and impact velocities necessary to attain filling factors observed in chondrites. Detailed analysis of collisions of chondrules with dust rims by \citet{GKDU17} and \citet{UGDU18} using granular mechanics simulation showed that the bouncing velocity increases with greater filling factors and thicker dust rims. 
Most modeling studies on chondrule rim growth and dust coagulation assume that collisions occur in the hit-and-stick regime \cite{HK18, OCT08}.

The assumption of hit-and-stick collisions, however, only works well for low collision
velocities. It fails for high relative velocities, which leads to restructuring or fragmentation due to excessive collision energy. Using smoothed particle hydrodynamics (SPH)  \Citet{MGP24} found that compaction during fragmentation and bouncing produces realistic dust grain porosities in protoplanetary discs. Detailed information on the collision and fragmentation processes of asteroids, planetesimals and larger bodies has been given by this method \cite{WGBB09, GMSK11}. However, models of the temporal evolution of
dust rims rely on information about the
outcome of individual cluster-cluster collisions. Here we focus on the results of high-energy collisions and characterize the extent of rim restructuring as a function of the collision energy.

The paper is organized as follows. Section \ref{Method} describes the granular-mechanics algorithm, the collision model, and the parameter section. Section \ref{Results} gives the results for the change in rim porosity and thickness as a function of the collision energy. The predicted rim growth for various levels of turbulence within the solar nebula is presented in Section \ref{ML Prediction}. Concluding remarks are given in Section \ref{Conclusions}.

\section{Method}  \label{Method}

\subsection{Granular-mechanics algorithm}  \label{s_alg}

We implemented our granular mechanics algorithm \cite{RU12_cpc} in the open source LIGGGHTS code \cite{LIGGGHTS}. LIGGGHTS is an extension of the popular molecular dynamics code LAMMPS \cite{LAMMPS} for the simulation of granular media. The acronym means `LAMMPS Improved for General Granular and Granular Heat Transfer Simulations'. LIGGGHTS provides high computational efficiency and is parallelized; it also offers advanced features, such as the use of complex geometries and the implementation of heat conduction between particles, which are, however, not used here.

Our code has been applied to simulate collisions of granular clusters \cite{RBBU12}, and the impact
of a hard sphere or aggregate of hard spheres on a granular bed representing a section of a chondrule dust rim \cite{RBU12, XCH19}. The algorithm has
been devised so as to model systems containing large
numbers ($>10^5$) of grains; therefore, the collision dynamics contains
simplifying assumptions. The validity of these assumptions has been tested as shown in \cite{RBBU12}.
In the following, we briefly describe the main features of our method.

Our clusters consist of spherical grains, all of which possess the same material properties (elastic moduli, etc.), but have different radii $\Rgrain$. The grains only
interact if the distance of their centers $d<R_{grain1}+R_{grain2}$. As is common
in granular mechanics, the length $\delta = (R_{grain1}+R_{grain2}) - d$ is called the
grain \emph{overlap}, and interactions are nonzero only for $\delta>0$.

Forces between grains are classified as \emph{normal} and
\emph{tangential forces}. The normal force consists of repulsive and
adhesive contributions. The repulsive part \cite{PS05book},
\begin{equation}  \label{rep}
f_{\rm rep}=
\frac{4}{3} M
\sqrt{\Rred \delta}
(\delta+A v_n) ,
\end{equation}
consists of a Hertzian $\delta^{3/2}$ contribution, based on elastic
theory, and a dissipative part, describing a viscoelastic contact
\cite{BSHP96}. Here $\Rred= \Rgrain/2$ is the reduced radius,
$M=Y/2(1-\nu^2)$ is the reduced modulus, $Y$ Young's modulus, $\nu$
Poisson's ratio, $v_n$ is the velocity component in normal direction,
and $A$ is an empirical factor modeling dissipation. The adhesive part
of the normal force is taken to be proportional to the
specific surface energy
\cite{DMT75,Mau00,Blu06},
\begin{equation} \label{e_f_adh}
f_{\rm adh}= 4 \pi \Rred \gamma_{\rm eff} .
\end{equation}
Here $\gamma_{\rm eff}=2\gamma$ is the effective specific surface energy of the system, equal to twice the surface energy of a single grain.

Note that this value corresponds to the pull-off force of DMT
theory \cite{DMT75} needed to break a
contact. We simplify the complex dependence of the adhesive force \textemdash as given by DMT \cite{DMT75} or JKR theory \cite{JKR71,Joh85book} \textemdash by a constant value \cite{RU12_cpc}. In our approach, this value
equals the pull-off force of DMT theory, which is valid for small particles with high elastic moduli (high rigidity), such as in our system. JKR theory is better suited to systems consisting of large grains with a low stiffness.

The tangential forces consist of several friction
forces. Gliding friction,
\begin{equation} \label{e_sl}
f_{\rm slide}=\frac{1}{2}  G  \pi a^2 ,
\end{equation}
depends on the shear modulus $G$ and the radius $a=\sqrt{\delta \Rred}$
of the contact area \cite{BK99}. Rolling motion is decelerated by a torque \cite{DT97},
\begin{equation}  \label{e_ro}
D_r=2  f_{\rm adh} \xiyield .
\end{equation}
Here, $\xiyield$ is the distance which two grains can roll over each other without breaking their atomic contacts. Finally, also torsional motion is decelerated by a torque, whose strength is given by
\cite{DT97}
\begin{equation}  \label{e_tw}
D_t=\frac{1}{3} G  \frac{a^3}{\pi} .
\end{equation}

In the actual implementation, we supplement the velocity independent friction force, \qeq{e_sl}, with a velocity proportional contribution valid for small velocities. This has the effect that the abrupt jump in the sliding force, which occurs when the tangential velocity approaches zero and leads to oscillations in the sign of the friction force from
time step to time step, is smoothed. In detail, following \cite{HW86} and \cite{PS05book}, we write the tangential force $f_t$ as
  \begin{equation} \label{e_visc}
    f_t = -{\rm sgn}
    (v_t) \cdot \min \left\{ \eta_{\rm tang} v_t ,
    f_{\rm slide} \right\} .
  \end{equation}
The value of the damping constant $\eta_{\rm tang}$ is chosen such that
the above-mentioned oscillations do not occur. The choice of $\eta_{\rm
tang}$ thus depends on the adopted time step, $\Delta t$, and is best
determined by a test simulation. We found a value of $\eta_{\rm
tang}=0.1 m/\Delta t$ appropriate.
As a consequence, grain contact is
stabilized. For the other friction forces we proceed analogously. For
details see Ref.\ \cite{RU12_cpc}.

There are two models to calculate the frictional force between two monomers, Hertzian and Hookean. We use the Hertzian model, in which the normal push-back force for two overlapping monomers is proportional to the area of overlap of the two monomers. The overlap depends on the monomer sizes captured in the radii-dependent prefactors. In the Hookean model,  the normal push-back force is a function of the overlap distance and is independent of monomer size, and therefore is not suitable for polydisperse monomers.

Our simulations have been performed with the well-documented
molecular-dynamics package LIGGGHTS \cite{LIGGGHTS}, with the addition of
the above features.

\subsection{Modeling collisions}
\begingroup
As described above we used LIGGGHTS to simulate collisions between a layer of dust on a chondrule surface and dust particles (either monomers or aggregates). For computational expediency, we assume that the chondrule accretes dust isotropically \cite{BHP11}, and restrict our study to a small patch on the chondrule surface \cite{XCH19, Xiang21}. We are interested in the collisional outcomes that affect the rim structure. The collision energy, a function of the particles' relative velocity, is a critical factor in determining the outcome: incoming dust particles can stick at the point of contact, cause restructuring within the dust rim, bounce, or shatter the rim. We aim to evaluate the extent of the regions affected within the rim and the change in rim porosity as a function of the collision energy. 

To explore these dynamics, we built a library of dust piles with parameters assuming turbulent-driven collisions between dust and chondrule cores \cite{XCH19}. The cylindrical dust piles have porosities ranging from 0.73 to 0.98 and thicknesses from 10 to 118 $\mu$m, with a radial extent of 50 $\mu$m. The potentially colliding dust grains are selected from a population of 1) spherical dust monomers with radii ranging from 0.5 to 10 $\mu$m or 2) small aggregates of these dust monomers, with equivalent radii from 0.47 to 51 $\mu$m. The equivalent radius, $R_{\sigma}$, is defined by equating the projected cross-sectional area of an aggregate averaged over many orientations to the area of a circle, $A_{\sigma} = \pi R_{\sigma}^2$. In each simulation, a pre-built rim and a colliding dust particle are randomly selected from the respective libraries. A random velocity, ranging from 0.01 to 100 \(\text{m/s}\), is assigned along with a random incident angle between \(0^\circ\) and \(30^\circ\). A drag force is applied to monomers that move beyond the radial extent of the pile during fragmentation or restructuring to simulate the resistance exerted by the unmodeled portion of the rim surrounding the pile. The LIGGGHTS simulations are stopped after 5~$\mu$s. This total simulated time is long enough for the collisions to produce the final shapes and filling factors, and most of the energy has been dissipated.   Snapshots from a sample collision are shown in Fig. \ref{fig:collision}.

\begin{figure*}[hbt!]
\centering
\includegraphics[width = 1\textwidth]{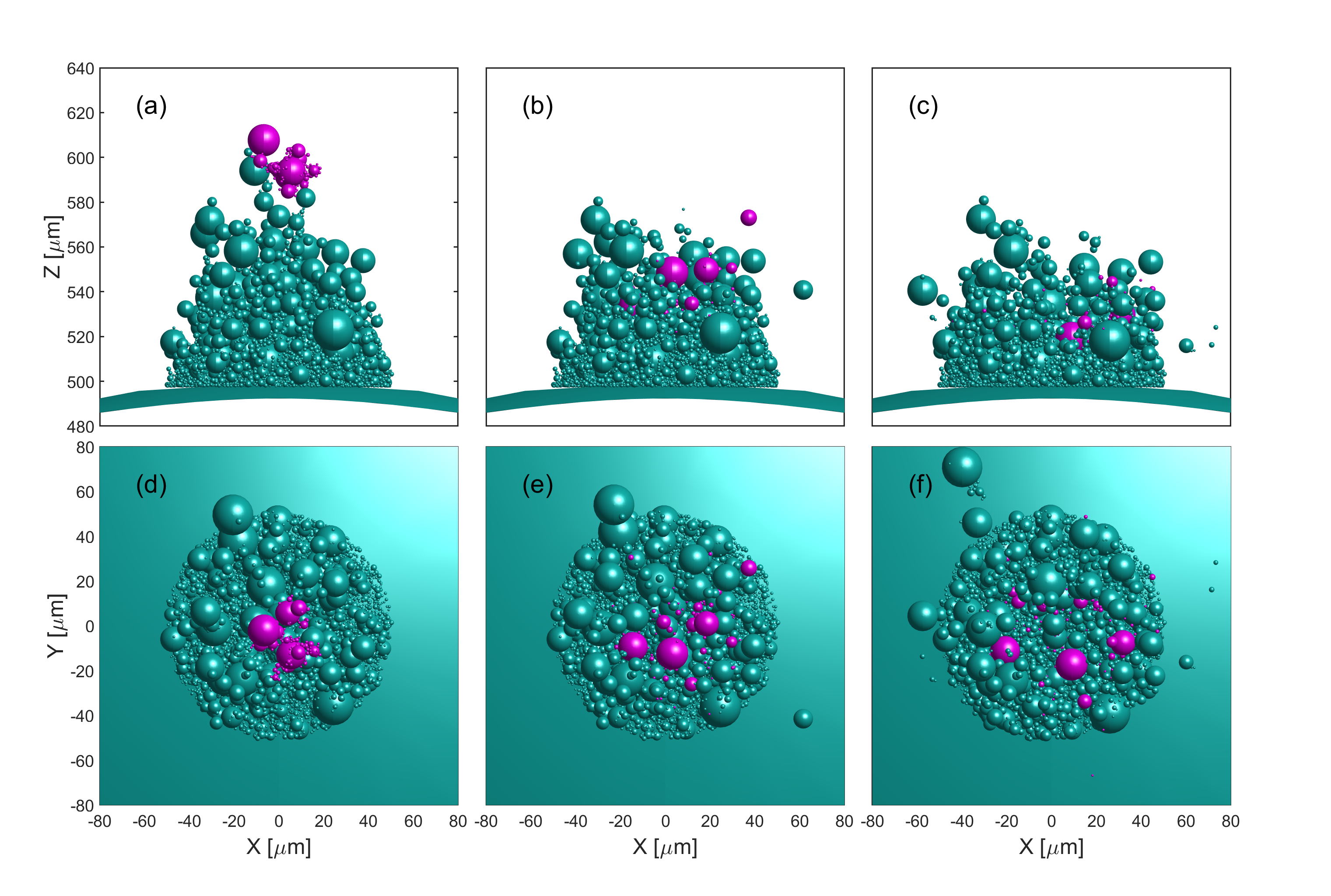}
\caption{Snapshots of a sample collision with views from the side (a-c) and from above (d-f). Panels (a) and (d) show the positions of the particles when the incoming particle (in purple) first touches the rim.  Panels (b) and (e) are mid-collision and panels (c) and (f) show the end result.  The incoming aggregate has an equivalent radius of 13.3 $\mu$m and collision energy KE of $1.53 \times 10^{-8}$ J.}
\label{fig:collision}
\end{figure*}

We characterize the collision outcomes as either causing restructuring or fragmentation. In the case of restructuring, individual monomers within the pile roll or slide on the surface of other monomers until they reach a new stable configuration as the energy is dissipated.  We are interested in both the total number of monomers that are displaced as well as the range of restructuring, defined as the maximum distance from the impact point. A monomer is considered to have been ``displaced" if its position relative to the chondrule core changes by more than 1\%. Given that the radius of the chondrule core is 500 $\mu$m, this distance is on the order of a few micrometers. A particle is classified as ``lost" from the pile if it is no longer in contact with either the chondrule surface or any other dust grains within the rim. The displaced and lost particles from the collision shown in Fig. \ref{fig:collision} are shown in Fig. \ref{fig:collision_result}a, and the final rim is shown in Fig. \ref{fig:collision_result}b.
\endgroup

\begin{figure*}[hbt!]
\centering
\includegraphics[width = 0.45\textwidth]{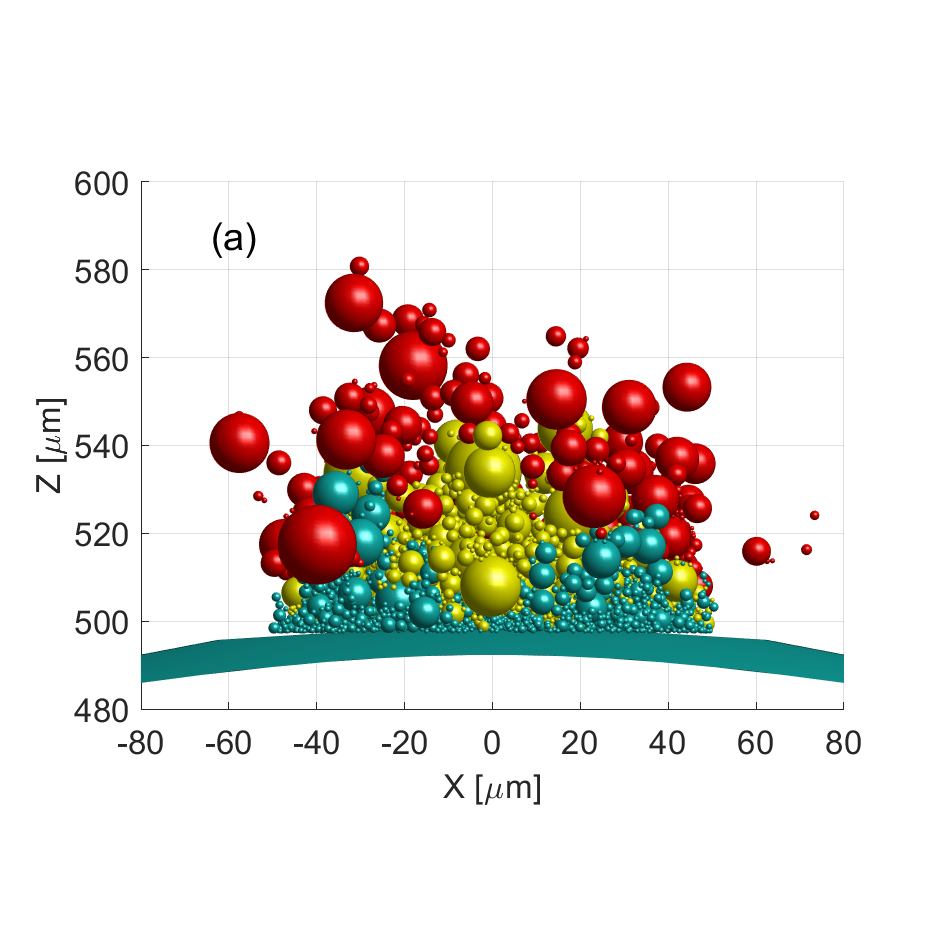}
\includegraphics[width = 0.45\textwidth]{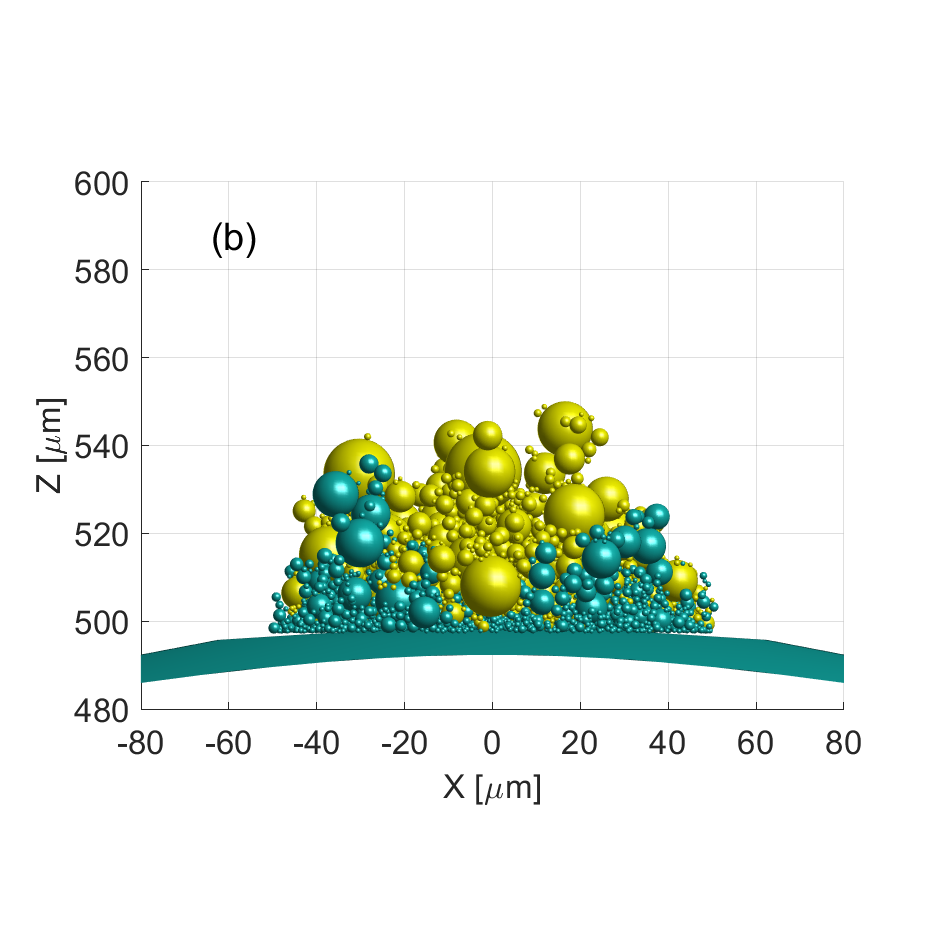}
\caption{Final result of the collision shown in Fig. \ref{fig:collision}. (a) Shows the monomers which are lost (red), displaced (yellow) and undisturbed (green).  (b) Same as (a) but with the displaced monomers removed.}
\label{fig:collision_result}
\end{figure*}

\subsection{Parameter selection}

Analysis of the chemical composition of chondrules finds that some chondrules must have formed in collisions between silicate-rich and ice-rich objects \cite{Mar16}.  As a simplifying assumption, we consider our grains to be made of a single type of material,
water ice, and the materials parameters are chosen as appropriate for this material:
Young's modulus $Y=70$ GPa, Poisson's ratio $\nu=0.31$, shear modulus
$G=Y/[2(1+\nu)]$, and 
surface energy $\gamma=0.19$ J/m$^{2}$.
The density of the ice
grains is
$\rho= 1$ g cm$^{-3}$. 
The dissipation constant $A$ has been fitted to the experimentally measured coefficient of restitution of ice grains. 

We note that the surface energy not only enters the equation for the adhesion force, but also rolling friction. In addition, due to the higher resulting equilibrium overlap, a higher dissipation in glide and torsional friction results.  We also note that changes in adhesion can cause changes in the size distributions of a collection of colliding grains \cite{GAS*12}.

\section{Results} \label{Results}

\begin{figure}[hbt!]
\centering\includegraphics[width = 0.48\textwidth]{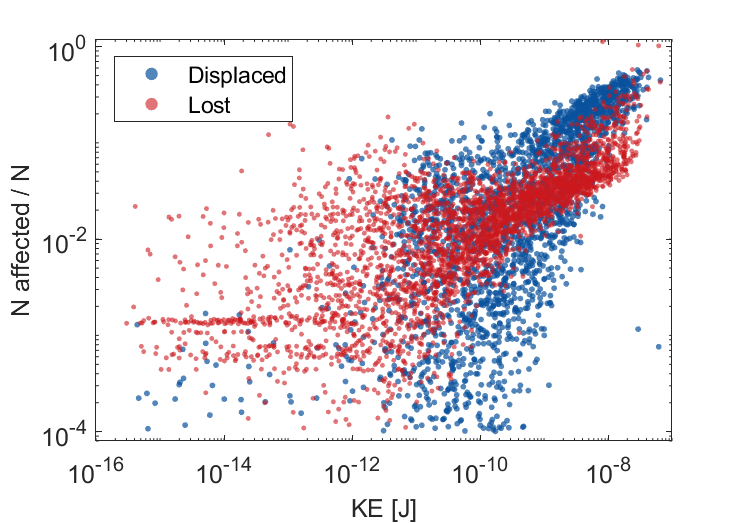}
\caption{{The ratio of monomers that are displaced within or lost from the rim to the total number of monomers in the rim, as a function of the collision energy.}}
\label{fig:N_affected_monomers}
\end{figure}

\begin{figure}[hbt!]
\centering\includegraphics[width = 0.48\textwidth]{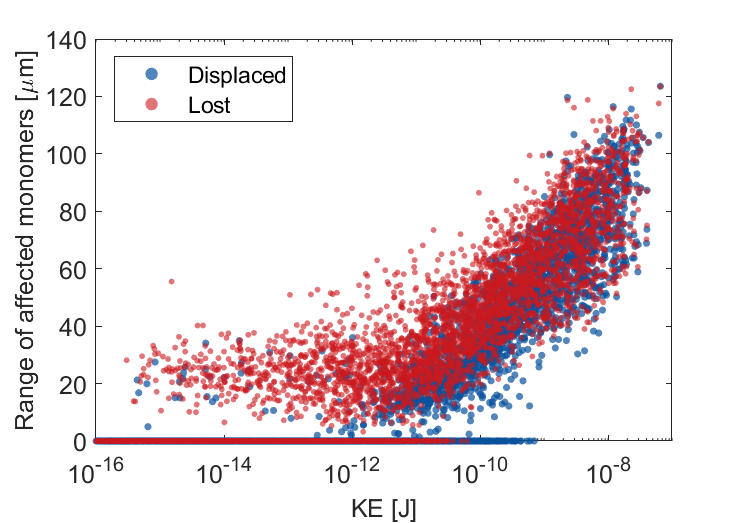}
\caption{Maximum distance from the point of impact for which monomers are displaced or lost as a function of the collision energy.  }
\label{fig:Range_affected_monomers}
\end{figure}

The data below show the results of $\sim$6000 individual collision events. The number of monomers displaced within or lost from a rim is shown in  Fig.~\ref{fig:N_affected_monomers} as a function of the collision energy. The range of the monomers affected (from the point of impact) is shown in Fig.~\ref{fig:Range_affected_monomers}. Note that in almost every collision where restructuring occurs, there are some monomers which are displaced and some which are knocked free and lost from the rim. For collision energies exceeding {$5 \times 10^{-9}$ J}, the average number of displaced monomers exceeds the number of monomers lost from the rim. It is interesting that at low collision energy, restructuring within the rim is more likely to be from monomers lost from the rim, rather than from displaced monomers. The range over which monomers are affected by restructuring is similar for both the displaced and the lost monomers, though the range of influence for lost monomers tends to be slightly greater. (In these simulations, the maximum range of restructuring is a three-dimensional distance which is limited by the diameter ($\sim$100$\mu$m) and the height (10 - 120 $\mu$m) of the pile.) 

\begin{figure}[hbt!]
\centering\includegraphics[width = 0.45\textwidth]{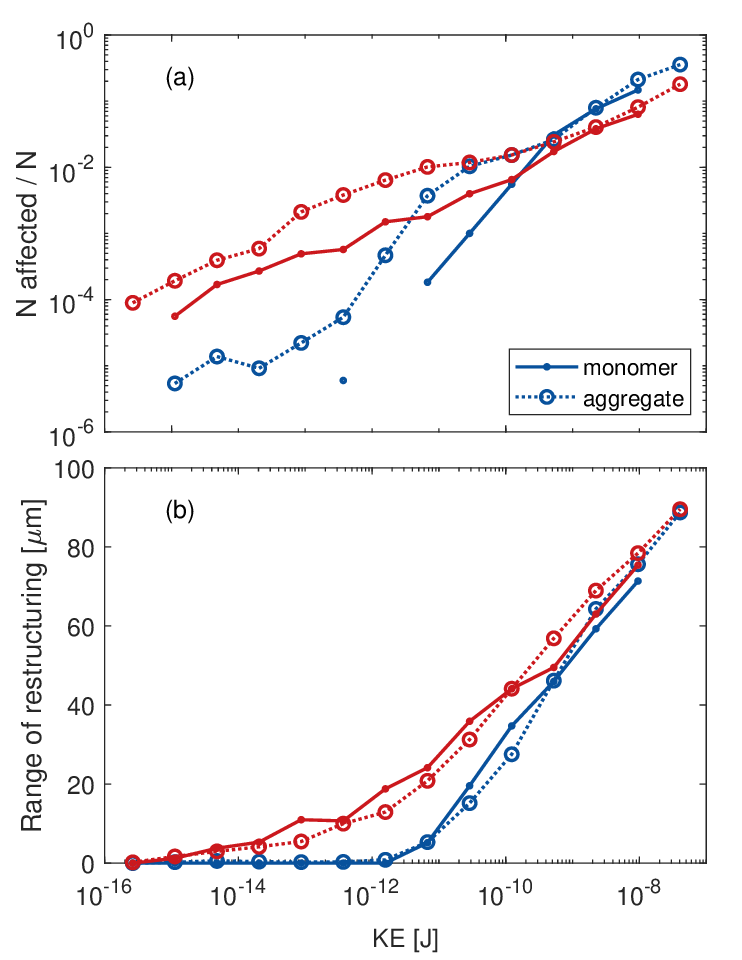}
\caption{a) Comparison of restructuring caused by collisions with monomers and aggregates. a) The number of monomers affected by restructuring and b) maximum distance from the impact point of the restructured monomers. The blue lines indicate monomers which are displaced and the red lines indicate monomers which are lost from the rim.}
\label{fig:monomer_vs_aggregate}
\end{figure}

Some insight is gained by comparing the results of collisions with single monomers to those of collisions with small aggregates. In Fig.~\ref{fig:monomer_vs_aggregate} the collision results have been binned based on the kinetic energy of the incoming particle, and the average number of displaced (blue lines) or lost monomers (red lines) is calculated for each bin. It appears that in lower energy collisions ($KE < \approx 10^{-10}$ J), colliding aggregates cause a greater number of particles to be displaced or lost from the pile (Fig.~\ref{fig:monomer_vs_aggregate}a), but affect a slightly smaller range (Fig.~\ref{fig:monomer_vs_aggregate}b). However, as the collision energy increases ($KE > \approx 10^{-9}$ J), aggregates affect a broader range of particles compared to monomers, largely due to their greater size. Note, however, that the highest energy collisions affected almost all of the monomers in the pile, which imposes an upper limit for the range of restructuring.  

\begin{figure}[hbt!]
\centering\includegraphics[width = 0.45\textwidth]{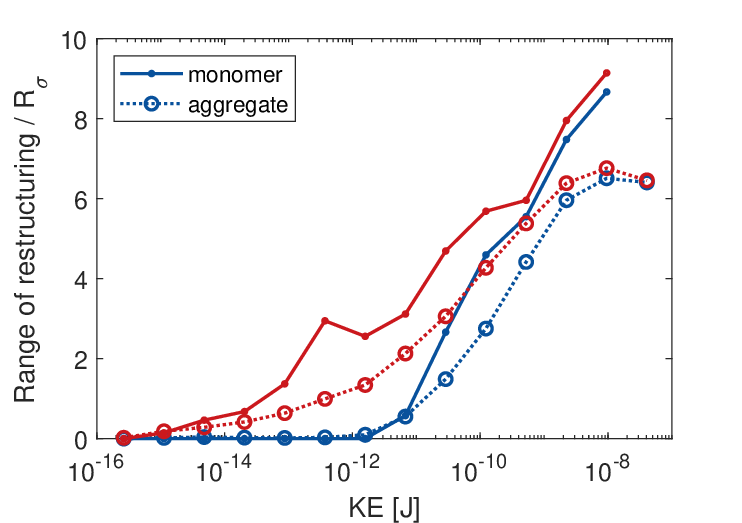}
\caption{Colliding monomers (solid lines, dots) cause restructuring of rim particles over a larger range relative to their size than do aggregates (dashed lines, open circles).  The range of rim particles affected in a collision is normalized by the  reduced radius of the colliding particle, $R_{sigma}$.  The blue lines indicate rim particles which are displaced, while the red line indicate particles with are lost from the rim.}
\label{fig:normalized_range}
\end{figure}

To compare the range of influence to the size of the impacting particle, we normalize the range by the equivalent radius, $R_{\sigma}$, as shown in Fig.~\ref{fig:normalized_range}. It is evident that monomer collisions have a greater range of influence relative to their size. A colliding monomer delivers energy concentrated in a small region, which means all of the energy is delivered to a few monomers in the rim. There is less energy loss as the collision energy is transmitted to adjacent particles. A large aggregate impacts a larger region of the rim, and therefore more rim particles, but less energy is transmitted to each rim particle and the effect is quickly damped. The leveling-off in the range of restructuring caused by aggregates in high energy collisions is due to the finite size of the pile.

\begin{figure}[hbt!]
\centering\includegraphics[width = 0.45\textwidth]{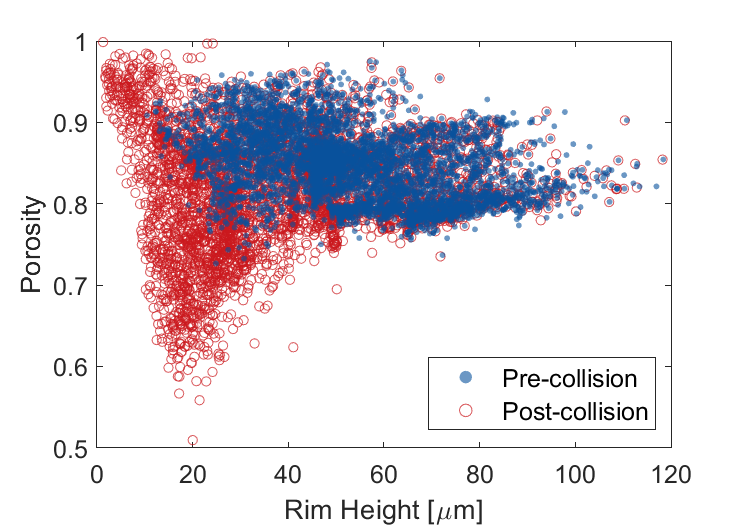}
\caption{Physical characteristics of the rims pre- and post-collision.  In general, rims become more compact with less porosity and a smaller thickness. }
\label{fig:pre_and_post_rim}
\end{figure}

\begin{figure}[hbt!]
\centering\includegraphics[width = 0.45\textwidth]{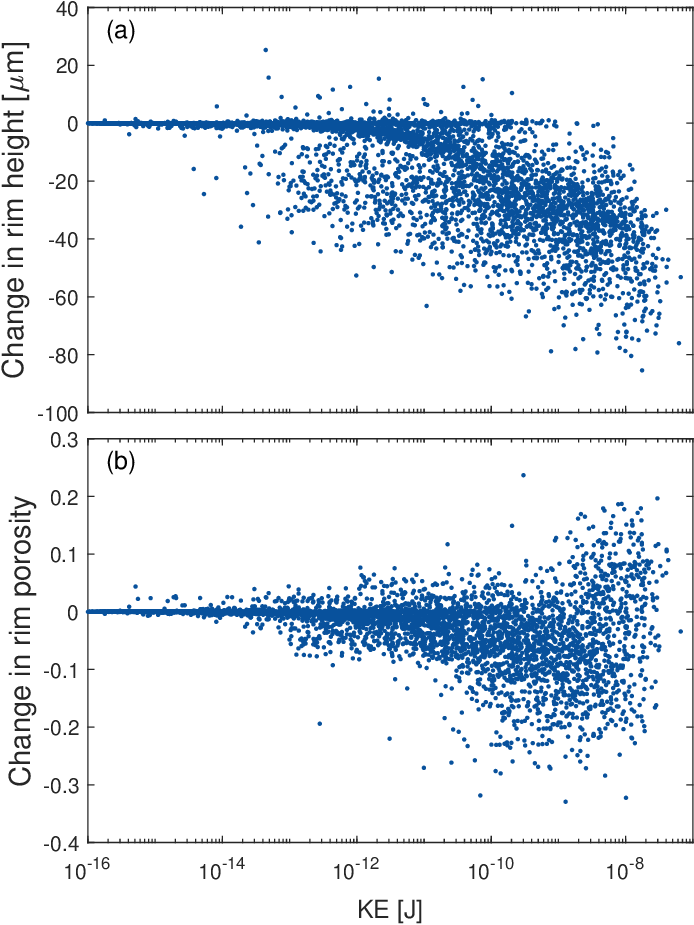}
\caption{Change in the physical characteristics of a chondrule rim as a function of collision energy. a) Most collisions cause a decrease in rim height. b) Most collisions result in a decreased porosity, but higher energy collisions are more likely to lead to an increase in porosity. }
\label{fig:changes_in_rim}
\end{figure}

The thickness and porosity of the rims pre- and post-collision are shown in Fig. ~\ref{fig:pre_and_post_rim}.  In general, the rims are smaller and less porous after collisions.  The change in rim thickness and porosity are shown in Fig.~\ref{fig:changes_in_rim} as a function of the kinetic energy. We define the height of the rim as the distance from the surface of chondrule core encompassing 95\% of the total rim mass.  The porosity is a measure of the open space within the rim, calculated as a fraction of the total volume of the rim. A cylindrical region with a radius half of that of the pile is used in the calculation or porosity in order to avoid edge effects. In most cases, energetic collisions reduce rim thickness, as shown in Fig.~\ref{fig:changes_in_rim}a. However, rim porosity can either increase or decrease as a result of high-energy collisions. The rim porosity decreases when more monomers are displaced within the rim, filling in the voids.  Losing monomers from the rim increases the porosity.  As collision energies exceed $10^{-10}$ J, a greater percentage of collisions lead to an increase in rim porosity due to the loss of rim material.

\begin{figure}[hbt!]
\centering\includegraphics[width = 0.45\textwidth]{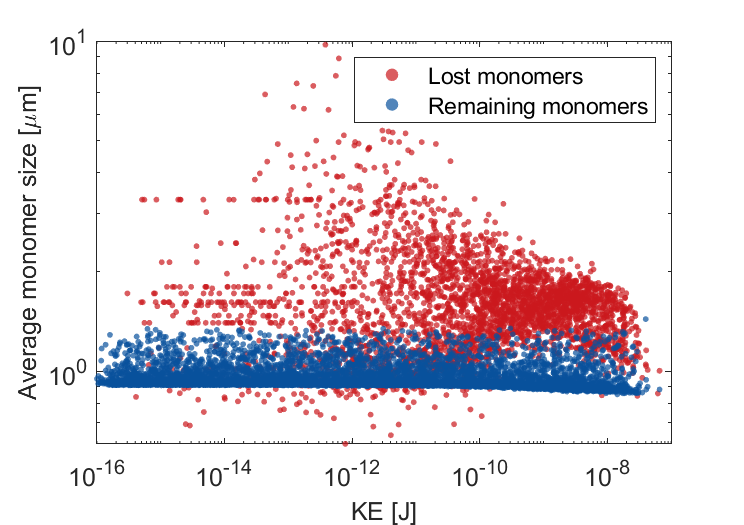}
\caption{The average size of the monomers which are lost (red) and remain attached (blue) during each collision event.}
\label{fig:avg_size_monomers}
\end{figure}

\begin{figure}[hbt!]
\centering\includegraphics[width = 0.45\textwidth]{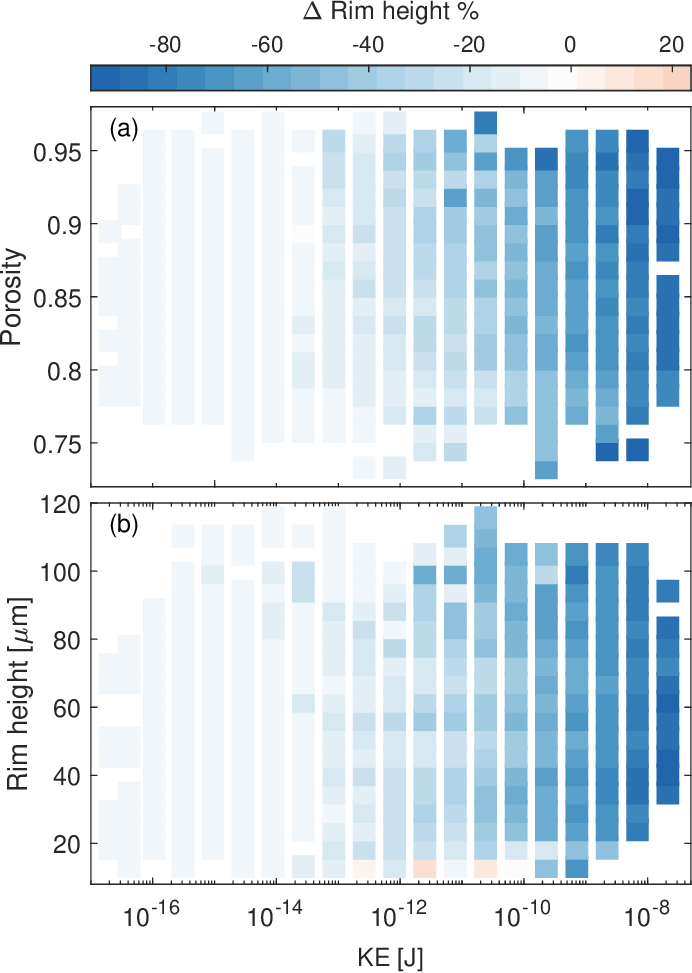}
\caption{Percent change in the thickness of the rim as a function of the kinetic energy of the collision and a) the original porosity of the rim and b) the original thickness of the rim.}
\label{fig:changes_in_rim_thickness}
\end{figure}

\begin{figure}[hbt!]
\centering\includegraphics[width = 0.45\textwidth]{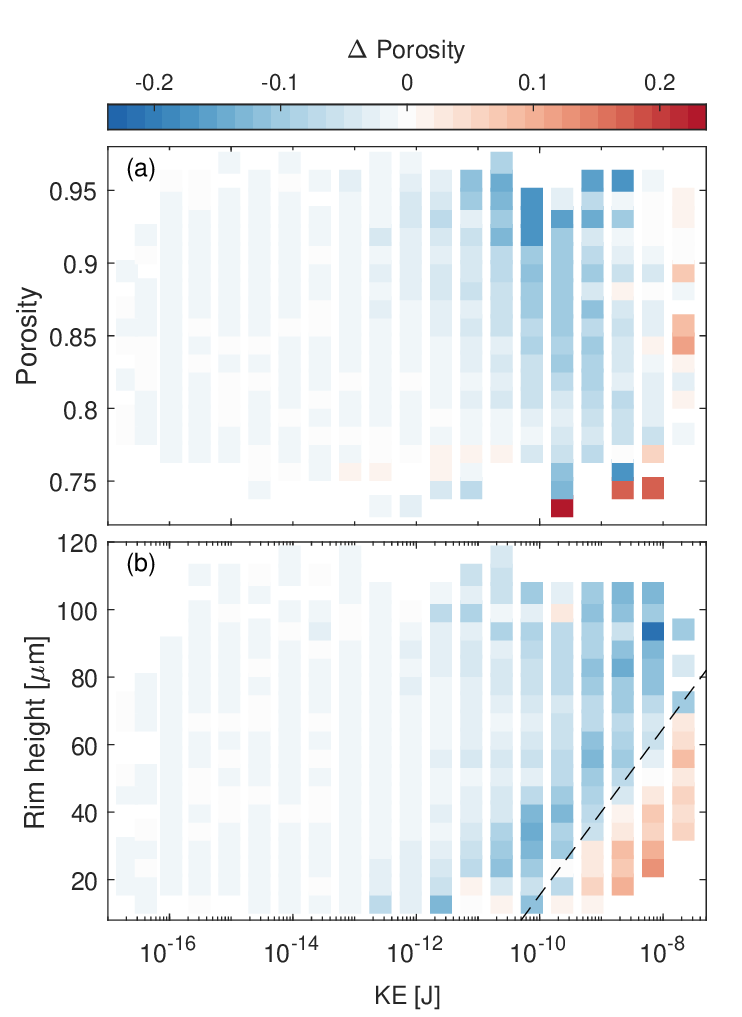}
\caption{Change in the porosity of the rim as a function of the kinetic energy of the collision and a) the original porosity of the rim and b) the original thickness of the rim. The dashed lines indicate where the change transitions from decreasing to increasing.}
\label{fig:changes_in_rim_porosity}
\end{figure}

There has been some speculation that porous chondrule rims may be able to absorb more collision energy and allow chondrules to stick together \cite{OCT08}. 
 The collision results were analyzed to see if thicker, more porous rims affected the rim restructuring. 
 Figures \ref{fig:changes_in_rim_thickness} and \ref{fig:changes_in_rim_porosity} illustrate how the kinetic energy of the dust particle and the original thickness and porosity of the rim affect the collision outcome, as characterized by the change in rim thickness and porosity. The data from the collisions are divided into logarithmically-spaced bins in kinetic energy and linearly-spaced bins in original rim thickness or porosity. The average changes in rim thickness or porosity in each bin are indicated by the color gradient. As expected, higher kinetic energy typically results in a greater reduction in rim thickness as more particles are restructured and knocked free. The most energetic collisions modeled ($KE > 2 \times 10^{-8} J$) lead to more than 80\%  of the rim height being lost, as shown in Fig. \ref{fig:changes_in_rim_thickness}. Given that the number of restructured monomers exceeds the number of lost monomers (Fig. \ref{fig:N_affected_monomers}), for there to be this much loss in thickness the large monomers must be preferentially lost. The small monomers remain attached to the chondrule surface due to their larger binding energy. This is indeed true, with the average size of the lost monomers with the trend becoming more slightly more pronounced for collision energies $KE > 10^{-10}$ J, as shown in Fig. \ref{fig:avg_size_monomers}. 
 
From Fig. \ref{fig:changes_in_rim_porosity}, it is evident that there is very little change in the porosity of the rim for collisions with $KE < 10^{-12}$ J. Collisions up to this threshold energy result in a slight decrease in the porosity of the rim as incoming particles fill in the void spaces.  As the energy increases above $10^{-12}$ J, restructuring within the rim  contributes to a more compact structure with a loss of porosity. Rims with porosity greater than 0.9 become more compact.  Accompanied with the decrease in rim height in Fig. \ref{fig:changes_in_rim_thickness}a, this indicates that the rims are being crushed.  The high-energy collisions which result in an increase in porosity also are accompanied by a loss in rim height, which indicates that in this case the majority of rim material is lost.  This is associated with thinner rims, as shown in Fig. \ref{fig:changes_in_rim_porosity}b. The region below the dashed line in Fig. \ref{fig:changes_in_rim_porosity}b indicates the region where particle ejection during fragmentation causes an increase in porosity. Just above this line, the decrease in porosity is most pronounced.  It is evident that compact and thin rims are particularly prone to increased porosity in high-energy collisions, as their limited capacity for energy dissipation through restructuring makes them more susceptible to fragmentation. 
 
\begin{figure}[hbt!]
\centering\includegraphics[width = 0.45\textwidth]{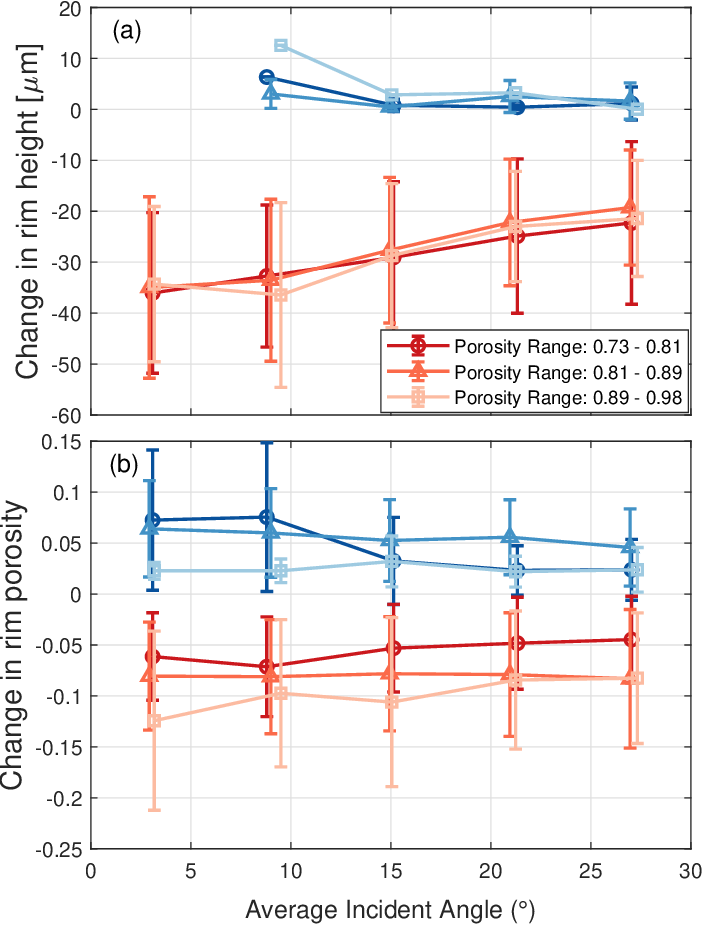}
\caption{Average change in (a) rim thickness and (b) rim porosity for high energy collisions (KE $> 10^{-11}$ J) as a function of the incident angle. The blue lines correspond to increased thickness or porosity and the red lines correspond to decreased  thickness or porosity. The data is grouped by the original rim porosity, where darker shades represent lower porosity and lighter shades represent greater porosity.}
\label{fig:incident_angle}
\end{figure}

 The changes in rim porosity and rim height are influenced in part by the incident angle shown in Fig. \ref{fig:incident_angle} for high energy collisions (KE $> 10^{-11}$ J). Of the 2337 collisions in this energy range, only 3\% resulted in increases in the rim height, while 31\% resulted in increased porosity.   Head-on collisions ($\theta < 5^{o}$) result in the greatest amount of crushing, resulting in reduced rim height. Impacts with a greater angle of incidence result in less crushing. The trend for the decrease in rim thickness is almost linear with respect to the incident angle and independent of the porosity, \ref{fig:incident_angle}a. Most collisions also result in a decrease in porosity as the rim is crushed, with head-on collisions leading to the greatest compaction (red lines, Fig. \ref{fig:incident_angle}b.) Here, porosity does play a role as the most porous rims have the greatest loss of porosity. This is to be expected as the fewer inter-particle connections in porous rims make monomers more likely to slide and restructure.  The most porous rims are also less likely to have an increase in porosity, especially in head-on collisions, as incoming particles fill in the open pore spaces.  

\section{Predicted Rim Growth} \label{ML Prediction}
A machine learning model was developed to predict changes in rim mass after collisions, based on the physical properties of the dust pile and particles, as well as the relative velocity and incident angle. The best results were obtained by splitting the data into two groups based on whether the rim mass increases or decreases after a collision and training separate models for each scenario. A classification model was first trained to predict whether the rim mass increases or decreases, with an accuracy of 88\%. To address the class imbalance that could bias the model toward the more frequent rim mass increase cases, the data was balanced before training.

For each scenario, the dataset was split into training and testing sets, with 80\% allocated for training and 20\% for testing. Various machine learning and deep learning models were evaluated, and the random forest model demonstrated the best performance for predicting mass changes, with hyperparameters optimized through a grid search. In cases where the rim mass increases, the model achieved a Mean Absolute Error (MAE) of 0.5 ng and a Root Mean Squared Error (RMSE) of 1.1 ng. For cases where the rim mass decreases, the MAE and RMSE were 2.2 ng and 3.4 ng, respectively. The maximum observed mass changes in the test data were 20.8 ng for increases in mass and 80.6 ng for decreases in mass.

The trained models were used to simulate the dust rim growth under different turbulence conditions. The simulations started with a bare chondrule surface, with a chondrule radius of $R_{ch}$ = 500 $\mu$m. The initial dust particle library consists of particles with equivalent radii ranging from 0.47 to 51 $\mu$m, following a power law size distribution $n(r)dr \propto r^{-3.5}dr$ where $n(r)dr$ is the number of particles in the size interval $(r, r+dr)$ \cite{Mathis77}. A Monte Carlo method was used to select an incoming particle from the dust population based on the collision rates between the rimmed chondrule with radius $R_{ch}$ and dust particle with radius $r_d$ \cite{Ormel07, OCT08}.

\begin{equation}  \label{collision_rate}
C_{ch,d} = \sigma_{ch,d}\Delta v_{ch,d}/V.
\end{equation}

\begin{figure}[hbt!]
\centering\includegraphics[width = 0.45\textwidth]{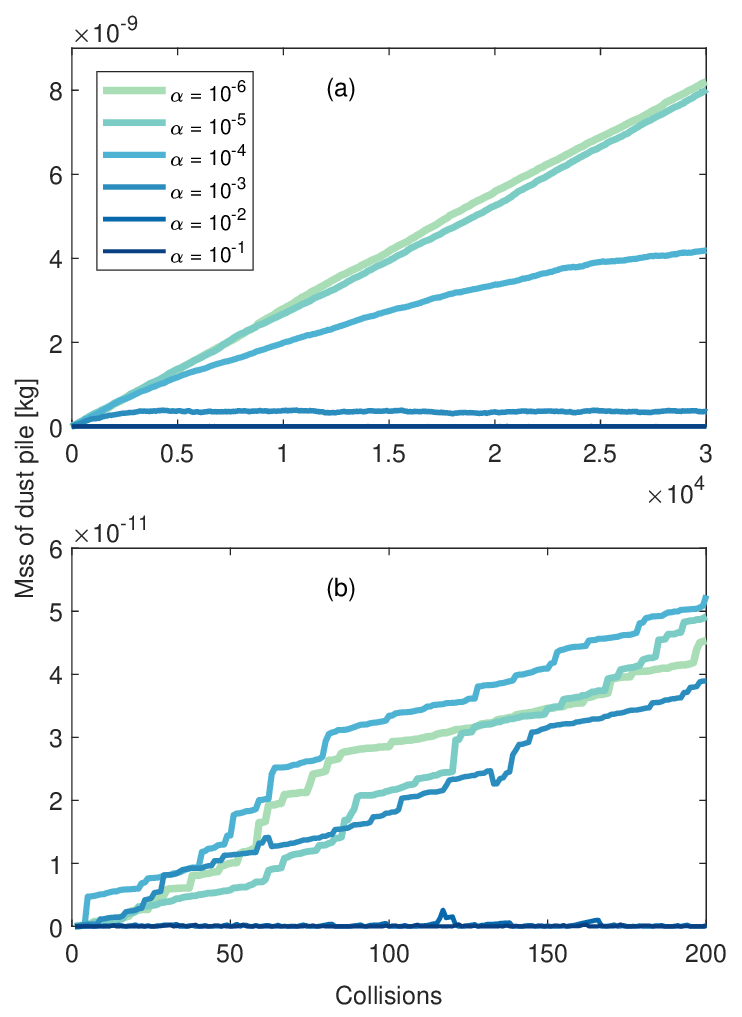}
\caption{a) Predicted growth of chondrule rims as a function of number of collisions. b) Rim growth during the first 150 collisions.}
\label{fig:predicted_rim_growth}
\end{figure}

The collision rate depends on the collision cross section $\sigma_{ch,d} = \pi (R_{ch} + r_d)^2$, the relative velocity of the dust with respect to the chondrule $\Delta v_{ch,d}$, and the volume $V$ of the simulated region, set by the density of the dust in the region. The relative velocity between the chondrule and the dust grains is set assuming that the dust is coupled to the turbulent eddies in the gas in the protoplanetary disk \cite{OCT08},

\begin{equation} \label{rel_vel}
v_T = v_g Re^{1/4}(St_1 - St_2)
\end{equation}
where $v_g$ is the gas speed and $Re$ is the Reynolds numbers, defined as the ratio of the turbulent viscosity, $\nu_{T}=\alpha c_{g}^{2}/\Omega$, to the molecular viscosity of gas, $\nu_{m}=c_{g}\lambda/2$ \cite{Cuzzi93}, with $\alpha$ the turbulence strength, $c_{g}$ the gas thermal speed, $\Omega$ the local Keplerian angular speed, and $\lambda$ the mean free path. $\mathrm{St}_i=\tau_i/t_L$ are the Stokes numbers of the two particles, with $\tau_i=\frac{3}{4c_{g}\rho _{g}}\frac{m_1}{\pi a_i^{2}}$ the stopping time of dust particle ($\rho_{g}$ is the gas density; $m_1$ and $a_1$ are the mass and equivalent radius of the particle), and $t_{L}=1/\Omega$ the turn-over time of the largest eddy. Given the mass and relative velocity of the particle selected, the kinetic energy of the collision was used to predict the change in the rim morphology.  The results are shown in Fig. \ref{fig:predicted_rim_growth} for each turbulence level.

In general, weaker turbulence promotes greater rim growth, due to fewer destructive collisions. Chondrule rims are able to continue to grow in weak turbulence, $\alpha \leq 10^{-4}$. At a turbulence level of $\alpha = 10^{-3}$, the rim undergoes periodic distruction, halting further growth, with the rim mass fluctuating around a small non-zero value. For turbulence level $\alpha \geq 10^{-2}$, the rims are rapidly eroded and no rim growth is able to occur.

\begin{figure}[hbt!]
\centering\includegraphics[width = 0.45\textwidth]{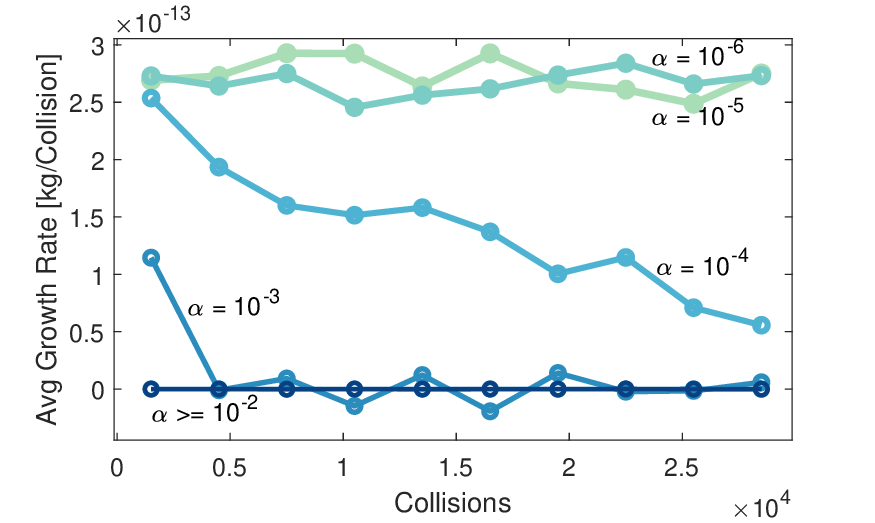}
\caption{Average rate of growth of the chondrule rim mass. The growth rate is calculated from the predicted rim mass shown in Fig. \ref{fig:predicted_rim_growth}  for turbulence levels ranging from $\alpha = 10^{-6} - 10^{-1}$.}
\label{fig:avg_growth_rate}
\end{figure}

Steady growth rates for $\alpha \leq 10^{-5}$ are evident in the approximately constant slope of the curves in Fig. \ref{fig:predicted_rim_growth}a. The growth rate as a function of turbulence level $\alpha$ is summarized in Fig. \ref{fig:avg_growth_rate}.  The results from cumulative collisions are divided into ten bins and the average change in rim mass during each collision is calculated for each bin. The rim growth is relatively constant for $\alpha \leq 10^{-5}$, indicating steady and unabated rim growth. Note that the growth rate here is a function of the number of collisions, as we are primarily examining the impact of a given collision outcome on rim growth. If the growth rate were calculated as a function of time, instead, it would decrease over time due to the depletion of the dust population. Note also that although the lowest turbulence level $\alpha = 10^{-6}$ has the highest growth rate per collision, the growth rate over time is smaller than for $\alpha = 10^{-5}$, since the time between collisions is longer for the lower relative velocities. At mid-level turbulence ($\alpha = 10^{-3}, 10^{-4}$), the growth rate decreases as the rims grow. The stronger the turbulence, the lower the growth rate and the more rapidly it declines. This is caused by the increased mass and reduced porosity of the rimmed chondrule, which leads to higher relative velocities with incoming dust particles, resulting in more frequent and destructive collisions. Our results are consistent with previous findings that weak turbulence facilitates rim growth through efficient sticking collisions, whereas strong turbulence slows or hinders rim growth due to frequent or continual disruptions \cite{OCT08, Cuzzi2004}.

\section{Conclusions} \label{Conclusions}
We have extended the study of chondrule rim growth to consider the results of disruptive collisions which occur when the relative velocity between colliding grains exceeds the ``hit-and-stick" threshold of $\approx$ 10 cm/s. The results of collisions are most readily characterized by the kinetic energy of the collision, based on the mass of the impacting particle and the relative velocity between the colliding pair. Below a threshold of $KE = 10^{-12}$, which is equivalent to relative velocities ranging between 0.13 - 27 m/s for the particles used in this study, very little restructuring takes place. As the collision energy increases up to $KE \approx 10^{-10}$ J, ($V_{rel} = 1.3 - 270 m/s$), the constituent particles in the rim undergo restructuring which tends to compact the rim, decreasing its thickness and porosity. Collisions with impact energy $ > 10^{-8}$ J  tend to completely disrupt the rim, with most of the particles being knocked free from the rim and lost. 

The relative velocity between particles, and by extension the kinetic energy of the collision, scales with the turbulence level in the disk.  The detailed collision results were used to train machine learning models to predict the change in rim thickness and porosity as a result of a collision.  These predictions were then used in a Monte Carlo simulation of rim growth, where incoming particles were selected randomly based on their probability of collision with a rimmed chondrule. We found that rim growth can continue unabated in regions where turbulence is weak, with $\alpha \leq 10^{-5}$. Chondrule rims are eroded in regions with mid-level turbulence, $\alpha =  10^{-3}, 10^{-4}$.  Chondrules can't collect rims in regions where the turbulence level is greater than this.

The particles used in this study were water ice. The restructuring expected for other constituent materials, such as silicates, will be affected by the binding energy of the material. Since the binding energy between water ice grains is 3.8 times greater than that for silicate material, the threshold kinetic energies may be expected to increase by a similar factor.

\begin{acknowledgments}
We gratefully acknowledge the High Performance and Research Computing Services team at Baylor University for providing computing resources and expertise to complete this work. This work was funded by NSF Astronomy and Astrophysics Research grant 2008493. N.M gratefully acknowledges the financial support from the Simulation Science Center Clausthal/G{\"o}ttingen.
\end{acknowledgments}

\section{Author Contributions}
Author Contributions
C. Xiang: Conceptualization (equal); Data curation (equal);
Formal analysis (lead); Investigation (equal); Methodology (equal);
Software (equal); Visualization (equal); Writing 
(equal). N. Merkert: Conceptualization (supporting);
Investigation (supporting); Software (equal); Validation (supporting);
Writing  (equal). L. Matthews:
Conceptualization (equal); Data curation (equal); Formal analysis
(equal); Funding acquisition (equal); Methodology (equal); Project
administration (equal); Resources (equal);  Supervision
(equal); Validation (equal); Visualization (equal); Writing 
(equal). 
A Carballido: Funding acquisition (lead); Project
administration (equal);  Supervision (equal); Writing
\textemdash review \& editing (supporting). T. Hyde: Funding acquisition (equal); Project
administration (equal); Supervision (equal); Writing
\textemdash review \& editing (supporting).


\bibliography{newGranular}

\begin{thebibliography}{39}
\expandafter\ifx\csname natexlab\endcsname\relax\def\natexlab#1{#1}\fi
\expandafter\ifx\csname bibnamefont\endcsname\relax
  \def\bibnamefont#1{#1}\fi
\expandafter\ifx\csname bibfnamefont\endcsname\relax
  \def\bibfnamefont#1{#1}\fi
\expandafter\ifx\csname citenamefont\endcsname\relax
  \def\citenamefont#1{#1}\fi
\expandafter\ifx\csname url\endcsname\relax
  \def\url#1{\texttt{#1}}\fi
\expandafter\ifx\csname urlprefix\endcsname\relax\def\urlprefix{URL }\fi
\providecommand{\bibinfo}[2]{#2}
\providecommand{\eprint}[2][]{\url{#2}}

\bibitem[{\citenamefont{Blum}(2010)}]{Blu10}
\bibinfo{author}{\bibfnamefont{J.}~\bibnamefont{Blum}}, \bibinfo{journal}{Research in Astron. Astrophys.} \textbf{\bibinfo{volume}{10}}, \bibinfo{pages}{1199} (\bibinfo{year}{2010}).

\bibitem[{\citenamefont{Reipurth et~al.}(2007)\citenamefont{Reipurth, Jewitt, and Keil}}]{Reipurth2007protostars}
\bibinfo{author}{\bibfnamefont{B.}~\bibnamefont{Reipurth}}, \bibinfo{author}{\bibfnamefont{D.}~\bibnamefont{Jewitt}}, \bibnamefont{and} \bibinfo{author}{\bibfnamefont{K.}~\bibnamefont{Keil}}, \emph{\bibinfo{title}{Protostars and planets V}} (\bibinfo{publisher}{University of Arizona Press}, \bibinfo{year}{2007}).

\bibitem[{\citenamefont{Johnson et~al.}(2015)\citenamefont{Johnson, Minton, Melosh, and Zuber}}]{JMMZ15}
\bibinfo{author}{\bibfnamefont{B.~C.} \bibnamefont{Johnson}}, \bibinfo{author}{\bibfnamefont{D.~A.} \bibnamefont{Minton}}, \bibinfo{author}{\bibfnamefont{H.~J.} \bibnamefont{Melosh}}, \bibnamefont{and} \bibinfo{author}{\bibfnamefont{M.~T.} \bibnamefont{Zuber}}, \bibinfo{journal}{Nature} \textbf{\bibinfo{volume}{517}}, \bibinfo{pages}{339} (\bibinfo{year}{2015}), ISSN \bibinfo{issn}{1476-4687}.

\bibitem[{\citenamefont{Hasegawa et~al.}(2015)\citenamefont{Hasegawa, Wakita, Matsumoto, and Oshino}}]{HWMO16}
\bibinfo{author}{\bibfnamefont{Y.}~\bibnamefont{Hasegawa}}, \bibinfo{author}{\bibfnamefont{S.}~\bibnamefont{Wakita}}, \bibinfo{author}{\bibfnamefont{Y.}~\bibnamefont{Matsumoto}}, \bibnamefont{and} \bibinfo{author}{\bibfnamefont{S.}~\bibnamefont{Oshino}}, \bibinfo{journal}{The Astrophysical Journal} \textbf{\bibinfo{volume}{816}}, \bibinfo{pages}{8} (\bibinfo{year}{2015}).

\bibitem[{\citenamefont{Wakita et~al.}(2017)\citenamefont{Wakita, Matsumoto, Oshino, and Hasegawa}}]{WMOH17}
\bibinfo{author}{\bibfnamefont{S.}~\bibnamefont{Wakita}}, \bibinfo{author}{\bibfnamefont{Y.}~\bibnamefont{Matsumoto}}, \bibinfo{author}{\bibfnamefont{S.}~\bibnamefont{Oshino}}, \bibnamefont{and} \bibinfo{author}{\bibfnamefont{Y.}~\bibnamefont{Hasegawa}}, \bibinfo{journal}{The Astrophysical Journal} \textbf{\bibinfo{volume}{834}}, \bibinfo{pages}{125} (\bibinfo{year}{2017}).

\bibitem[{\citenamefont{Ormel et~al.}(2008)\citenamefont{Ormel, Cuzzi, and Tielens}}]{OCT08}
\bibinfo{author}{\bibfnamefont{C.~W.} \bibnamefont{Ormel}}, \bibinfo{author}{\bibfnamefont{J.~N.} \bibnamefont{Cuzzi}}, \bibnamefont{and} \bibinfo{author}{\bibfnamefont{A.~G. G.~M.} \bibnamefont{Tielens}}, \bibinfo{journal}{The Astrophysical Journal} \textbf{\bibinfo{volume}{679}}, \bibinfo{pages}{1588} (\bibinfo{year}{2008}).

\bibitem[{\citenamefont{Xiang et~al.}(2019)\citenamefont{Xiang, Carballido, Hanna, Matthews, and Hyde}}]{XCH19}
\bibinfo{author}{\bibfnamefont{C.}~\bibnamefont{Xiang}}, \bibinfo{author}{\bibfnamefont{A.}~\bibnamefont{Carballido}}, \bibinfo{author}{\bibfnamefont{R.}~\bibnamefont{Hanna}}, \bibinfo{author}{\bibfnamefont{L.}~\bibnamefont{Matthews}}, \bibnamefont{and} \bibinfo{author}{\bibfnamefont{T.}~\bibnamefont{Hyde}}, \bibinfo{journal}{Icarus} \textbf{\bibinfo{volume}{321}}, \bibinfo{pages}{99} (\bibinfo{year}{2019}), ISSN \bibinfo{issn}{0019-1035}.

\bibitem[{\citenamefont{Xiang et~al.}(2023)\citenamefont{Xiang, Carballido, Matthews, and Hyde}}]{XCMH23}
\bibinfo{author}{\bibfnamefont{C.}~\bibnamefont{Xiang}}, \bibinfo{author}{\bibfnamefont{A.}~\bibnamefont{Carballido}}, \bibinfo{author}{\bibfnamefont{L.~S.} \bibnamefont{Matthews}}, \bibnamefont{and} \bibinfo{author}{\bibfnamefont{T.~W.} \bibnamefont{Hyde}}, \bibinfo{journal}{The Astrophysical Journal} \textbf{\bibinfo{volume}{950}}, \bibinfo{pages}{11} (\bibinfo{year}{2023}).

\bibitem[{\citenamefont{Brearley}(1993)}]{Bre93}
\bibinfo{author}{\bibfnamefont{A.~J.} \bibnamefont{Brearley}}, \bibinfo{journal}{Geochimica et Cosmochimica Acta} \textbf{\bibinfo{volume}{57}}, \bibinfo{pages}{1521} (\bibinfo{year}{1993}), ISSN \bibinfo{issn}{0016-7037}.

\bibitem[{\citenamefont{Hanna and Ketcham}(2018)}]{HK18}
\bibinfo{author}{\bibfnamefont{R.~D.} \bibnamefont{Hanna}} \bibnamefont{and} \bibinfo{author}{\bibfnamefont{R.~A.} \bibnamefont{Ketcham}}, \bibinfo{journal}{Earth and Planetary Science Letters} \textbf{\bibinfo{volume}{481}}, \bibinfo{pages}{201} (\bibinfo{year}{2018}), ISSN \bibinfo{issn}{0012-821X}.

\bibitem[{\citenamefont{Bland et~al.}(2011)\citenamefont{Bland, Howard, Prior, Wheeler, Hough, and Dyl}}]{BHP11}
\bibinfo{author}{\bibfnamefont{P.~A.} \bibnamefont{Bland}}, \bibinfo{author}{\bibfnamefont{L.~E.} \bibnamefont{Howard}}, \bibinfo{author}{\bibfnamefont{D.~J.} \bibnamefont{Prior}}, \bibinfo{author}{\bibfnamefont{J.}~\bibnamefont{Wheeler}}, \bibinfo{author}{\bibfnamefont{R.~M.} \bibnamefont{Hough}}, \bibnamefont{and} \bibinfo{author}{\bibfnamefont{K.~A.} \bibnamefont{Dyl}}, \bibinfo{journal}{Nature Geoscience} \textbf{\bibinfo{volume}{4}}, \bibinfo{pages}{244} (\bibinfo{year}{2011}), ISSN \bibinfo{issn}{1752-0908}.

\bibitem[{\citenamefont{Beitz et~al.}(2013)\citenamefont{Beitz, G\"uttler, Nakamura, Tsuchiyama, and Blum}}]{BGN13}
\bibinfo{author}{\bibfnamefont{E.}~\bibnamefont{Beitz}}, \bibinfo{author}{\bibfnamefont{C.}~\bibnamefont{G\"uttler}}, \bibinfo{author}{\bibfnamefont{A.}~\bibnamefont{Nakamura}}, \bibinfo{author}{\bibfnamefont{A.}~\bibnamefont{Tsuchiyama}}, \bibnamefont{and} \bibinfo{author}{\bibfnamefont{J.}~\bibnamefont{Blum}}, \bibinfo{journal}{Icarus} \textbf{\bibinfo{volume}{225}}, \bibinfo{pages}{558} (\bibinfo{year}{2013}), ISSN \bibinfo{issn}{0019-1035}.

\bibitem[{\citenamefont{Gunkelmann et~al.}(2017)\citenamefont{Gunkelmann, Kataoka, Dullemond, and Urbassek}}]{GKDU17}
\bibinfo{author}{\bibfnamefont{N.}~\bibnamefont{Gunkelmann}}, \bibinfo{author}{\bibfnamefont{A.}~\bibnamefont{Kataoka}}, \bibinfo{author}{\bibfnamefont{C.~P.} \bibnamefont{Dullemond}}, \bibnamefont{and} \bibinfo{author}{\bibfnamefont{H.~M.} \bibnamefont{Urbassek}}, \bibinfo{journal}{A\&A} \textbf{\bibinfo{volume}{599}}, \bibinfo{pages}{L4} (\bibinfo{year}{2017}).

\bibitem[{\citenamefont{Umst{\"a}tter et~al.}(2018)\citenamefont{Umst{\"a}tter, Gunkelmann, Dullemond, and Urbassek}}]{UGDU18}
\bibinfo{author}{\bibfnamefont{P.}~\bibnamefont{Umst{\"a}tter}}, \bibinfo{author}{\bibfnamefont{N.}~\bibnamefont{Gunkelmann}}, \bibinfo{author}{\bibfnamefont{C.~P.} \bibnamefont{Dullemond}}, \bibnamefont{and} \bibinfo{author}{\bibfnamefont{H.~M.} \bibnamefont{Urbassek}}, \bibinfo{journal}{Monthly Notices of the Royal Astronomical Society} \textbf{\bibinfo{volume}{483}}, \bibinfo{pages}{4938} (\bibinfo{year}{2018}), ISSN \bibinfo{issn}{0035-8711}.

\bibitem[{\citenamefont{Michoulier et~al.}(2024)\citenamefont{Michoulier, Gonzalez, and Price}}]{MGP24}
\bibinfo{author}{\bibfnamefont{S.}~\bibnamefont{Michoulier}}, \bibinfo{author}{\bibfnamefont{J.-F.} \bibnamefont{Gonzalez}}, \bibnamefont{and} \bibinfo{author}{\bibfnamefont{D.~J.} \bibnamefont{Price}}, \emph{\bibinfo{title}{Compaction during fragmentation and bouncing produces realistic dust grain porosities in protoplanetary discs}} (\bibinfo{year}{2024}), \eprint{2406.15622}.

\bibitem[{\citenamefont{Weidling et~al.}(2009)\citenamefont{Weidling, Güttler, Blum, and Brauer}}]{WGBB09}
\bibinfo{author}{\bibfnamefont{R.}~\bibnamefont{Weidling}}, \bibinfo{author}{\bibfnamefont{C.}~\bibnamefont{Güttler}}, \bibinfo{author}{\bibfnamefont{J.}~\bibnamefont{Blum}}, \bibnamefont{and} \bibinfo{author}{\bibfnamefont{F.}~\bibnamefont{Brauer}}, \bibinfo{journal}{Astron. J.} \textbf{\bibinfo{volume}{696}}, \bibinfo{pages}{2036} (\bibinfo{year}{2009}).

\bibitem[{\citenamefont{Geretshauser et~al.}(2011)\citenamefont{Geretshauser, Meru, Speith, and Kley}}]{GMSK11}
\bibinfo{author}{\bibfnamefont{R.~J.} \bibnamefont{Geretshauser}}, \bibinfo{author}{\bibfnamefont{F.}~\bibnamefont{Meru}}, \bibinfo{author}{\bibfnamefont{R.}~\bibnamefont{Speith}}, \bibnamefont{and} \bibinfo{author}{\bibfnamefont{W.}~\bibnamefont{Kley}}, \bibinfo{journal}{Astron. \& Astrophys.} \textbf{\bibinfo{volume}{531}}, \bibinfo{pages}{A166} (\bibinfo{year}{2011}).

\bibitem[{\citenamefont{Ringl and Urbassek}(2012)}]{RU12_cpc}
\bibinfo{author}{\bibfnamefont{C.}~\bibnamefont{Ringl}} \bibnamefont{and} \bibinfo{author}{\bibfnamefont{H.~M.} \bibnamefont{Urbassek}}, \bibinfo{journal}{Computer Physics Communications} \textbf{\bibinfo{volume}{183}}, \bibinfo{pages}{986} (\bibinfo{year}{2012}).

\bibitem[{\citenamefont{Kloss et~al.}(2012)\citenamefont{Kloss, Goniva, Hager, Amberger, and Pirker}}]{LIGGGHTS}
\bibinfo{author}{\bibfnamefont{C.}~\bibnamefont{Kloss}}, \bibinfo{author}{\bibfnamefont{C.}~\bibnamefont{Goniva}}, \bibinfo{author}{\bibfnamefont{A.}~\bibnamefont{Hager}}, \bibinfo{author}{\bibfnamefont{S.}~\bibnamefont{Amberger}}, \bibnamefont{and} \bibinfo{author}{\bibfnamefont{S.}~\bibnamefont{Pirker}}, \bibinfo{journal}{Prog. Comput. Fluid Dy.} \textbf{\bibinfo{volume}{12}}, \bibinfo{pages}{140} (\bibinfo{year}{2012}).

\bibitem[{\citenamefont{LAMMPS}(2008)}]{LAMMPS}
\bibinfo{author}{\bibnamefont{LAMMPS}}, \emph{\bibinfo{title}{http://lammps.sandia.gov/}} (\bibinfo{year}{2008}).

\bibitem[{\citenamefont{Ringl et~al.}(2012{\natexlab{a}})\citenamefont{Ringl, Bringa, Bertoldi, and Urbassek}}]{RBBU12}
\bibinfo{author}{\bibfnamefont{C.}~\bibnamefont{Ringl}}, \bibinfo{author}{\bibfnamefont{E.~M.} \bibnamefont{Bringa}}, \bibinfo{author}{\bibfnamefont{D.~S.} \bibnamefont{Bertoldi}}, \bibnamefont{and} \bibinfo{author}{\bibfnamefont{H.~M.} \bibnamefont{Urbassek}}, \bibinfo{journal}{Astrophysical Journal} \textbf{\bibinfo{volume}{752}}, \bibinfo{pages}{151} (\bibinfo{year}{2012}{\natexlab{a}}).

\bibitem[{\citenamefont{Ringl et~al.}(2012{\natexlab{b}})\citenamefont{Ringl, Bringa, and Urbassek}}]{RBU12}
\bibinfo{author}{\bibfnamefont{C.}~\bibnamefont{Ringl}}, \bibinfo{author}{\bibfnamefont{E.~M.} \bibnamefont{Bringa}}, \bibnamefont{and} \bibinfo{author}{\bibfnamefont{H.~M.} \bibnamefont{Urbassek}}, \bibinfo{journal}{Phys. Rev. E} \textbf{\bibinfo{volume}{86}}, \bibinfo{pages}{061313} (\bibinfo{year}{2012}{\natexlab{b}}).

\bibitem[{\citenamefont{P\"oschel and Schwager}(2005)}]{PS05book}
\bibinfo{author}{\bibfnamefont{T.}~\bibnamefont{P\"oschel}} \bibnamefont{and} \bibinfo{author}{\bibfnamefont{T.}~\bibnamefont{Schwager}}, \emph{\bibinfo{title}{Computational granular dynamics: models and algorithms}} (\bibinfo{publisher}{Springer}, \bibinfo{year}{2005}).

\bibitem[{\citenamefont{Brilliantov et~al.}(1996)\citenamefont{Brilliantov, Spahn, Hertzsch, and P\"oschel}}]{BSHP96}
\bibinfo{author}{\bibfnamefont{N.~V.} \bibnamefont{Brilliantov}}, \bibinfo{author}{\bibfnamefont{F.}~\bibnamefont{Spahn}}, \bibinfo{author}{\bibfnamefont{J.-M.} \bibnamefont{Hertzsch}}, \bibnamefont{and} \bibinfo{author}{\bibfnamefont{T.}~\bibnamefont{P\"oschel}}, \bibinfo{journal}{Phys. Rev. E} \textbf{\bibinfo{volume}{53}}, \bibinfo{pages}{5382} (\bibinfo{year}{1996}).

\bibitem[{\citenamefont{Derjaguin et~al.}(1975)\citenamefont{Derjaguin, Muller, and Toporov}}]{DMT75}
\bibinfo{author}{\bibfnamefont{B.~V.} \bibnamefont{Derjaguin}}, \bibinfo{author}{\bibfnamefont{V.~M.} \bibnamefont{Muller}}, \bibnamefont{and} \bibinfo{author}{\bibfnamefont{Y.~P.} \bibnamefont{Toporov}}, \bibinfo{journal}{Journal of Colloid and Interface Science} \textbf{\bibinfo{volume}{53}}, \bibinfo{pages}{314} (\bibinfo{year}{1975}).

\bibitem[{\citenamefont{Maugis}(2000)}]{Mau00}
\bibinfo{author}{\bibfnamefont{D.}~\bibnamefont{Maugis}}, \emph{\bibinfo{title}{Contact, adhesion and rupture of elastic solids}} (\bibinfo{publisher}{Springer}, \bibinfo{address}{Berlin}, \bibinfo{year}{2000}).

\bibitem[{\citenamefont{Blum}(2006)}]{Blu06}
\bibinfo{author}{\bibfnamefont{J.}~\bibnamefont{Blum}}, \bibinfo{journal}{Advances in Physics} \textbf{\bibinfo{volume}{55}}, \bibinfo{pages}{881} (\bibinfo{year}{2006}).

\bibitem[{\citenamefont{Johnson et~al.}(1971)\citenamefont{Johnson, Kendall, and Roberts}}]{JKR71}
\bibinfo{author}{\bibfnamefont{K.~L.} \bibnamefont{Johnson}}, \bibinfo{author}{\bibfnamefont{K.}~\bibnamefont{Kendall}}, \bibnamefont{and} \bibinfo{author}{\bibfnamefont{A.~D.} \bibnamefont{Roberts}}, \bibinfo{journal}{Proceedings of the Royal Society of London. A. mathematical and physical sciences} \textbf{\bibinfo{volume}{324}}, \bibinfo{pages}{301} (\bibinfo{year}{1971}).

\bibitem[{\citenamefont{Johnson}(1985)}]{Joh85book}
\bibinfo{author}{\bibfnamefont{K.~L.} \bibnamefont{Johnson}}, \emph{\bibinfo{title}{Contact mechanics}} (\bibinfo{publisher}{Cambridge University Press}, \bibinfo{address}{Cambridge}, \bibinfo{year}{1985}).

\bibitem[{\citenamefont{Burnham and Kulik}(1999)}]{BK99}
\bibinfo{author}{\bibfnamefont{N.}~\bibnamefont{Burnham}} \bibnamefont{and} \bibinfo{author}{\bibfnamefont{A.~A.} \bibnamefont{Kulik}}, in \emph{\bibinfo{booktitle}{Handbook of Micro/Nano Tribology}}, edited by \bibinfo{editor}{\bibfnamefont{B.}~\bibnamefont{Bhushan}} (\bibinfo{publisher}{CRC Press}, \bibinfo{address}{Boca Raton}, \bibinfo{year}{1999}), chap.~\bibinfo{chapter}{5}, p. \bibinfo{pages}{247}, \bibinfo{edition}{2nd} ed.

\bibitem[{\citenamefont{Dominik and Tielens}(1997)}]{DT97}
\bibinfo{author}{\bibfnamefont{C.}~\bibnamefont{Dominik}} \bibnamefont{and} \bibinfo{author}{\bibfnamefont{A.~G. G.~M.} \bibnamefont{Tielens}}, \bibinfo{journal}{Astrophys. J.} \textbf{\bibinfo{volume}{480}}, \bibinfo{pages}{647} (\bibinfo{year}{1997}).

\bibitem[{\citenamefont{Haff and Werner}(1986)}]{HW86}
\bibinfo{author}{\bibfnamefont{P.~K.} \bibnamefont{Haff}} \bibnamefont{and} \bibinfo{author}{\bibfnamefont{B.~T.} \bibnamefont{Werner}}, \bibinfo{journal}{Powder Technology} \textbf{\bibinfo{volume}{48}}, \bibinfo{pages}{239} (\bibinfo{year}{1986}).

\bibitem[{\citenamefont{Xiang et~al.}(2021)\citenamefont{Xiang, Carballido, Matthews, and Hyde}}]{Xiang21}
\bibinfo{author}{\bibfnamefont{C.}~\bibnamefont{Xiang}}, \bibinfo{author}{\bibfnamefont{A.}~\bibnamefont{Carballido}}, \bibinfo{author}{\bibfnamefont{L.~S.} \bibnamefont{Matthews}}, \bibnamefont{and} \bibinfo{author}{\bibfnamefont{T.~W.} \bibnamefont{Hyde}}, \bibinfo{journal}{Icarus} \textbf{\bibinfo{volume}{354}}, \bibinfo{pages}{114053} (\bibinfo{year}{2021}).

\bibitem[{\citenamefont{Marrochhi et~al.}(2016)\citenamefont{Marrochhi, Chaussidon, Piani, and Librourel}}]{Mar16}
\bibinfo{author}{\bibfnamefont{Y.}~\bibnamefont{Marrochhi}}, \bibinfo{author}{\bibfnamefont{M.}~\bibnamefont{Chaussidon}}, \bibinfo{author}{\bibfnamefont{L.}~\bibnamefont{Piani}}, \bibnamefont{and} \bibinfo{author}{\bibfnamefont{G.}~\bibnamefont{Librourel}}, \bibinfo{journal}{Science Advances} \textbf{\bibinfo{volume}{2}}, \bibinfo{pages}{e1601001} (\bibinfo{year}{2016}).

\bibitem[{\citenamefont{Guimaraes et~al.}(2012)\citenamefont{Guimaraes, Albers, Spahn, Seiss, Vieira-Neto, and Brilliantov}}]{GAS*12}
\bibinfo{author}{\bibfnamefont{A.~H.~F.} \bibnamefont{Guimaraes}}, \bibinfo{author}{\bibfnamefont{N.}~\bibnamefont{Albers}}, \bibinfo{author}{\bibfnamefont{F.}~\bibnamefont{Spahn}}, \bibinfo{author}{\bibfnamefont{M.}~\bibnamefont{Seiss}}, \bibinfo{author}{\bibfnamefont{E.}~\bibnamefont{Vieira-Neto}}, \bibnamefont{and} \bibinfo{author}{\bibfnamefont{N.~V.} \bibnamefont{Brilliantov}}, \bibinfo{journal}{Icarus} \textbf{\bibinfo{volume}{220}}, \bibinfo{pages}{660} (\bibinfo{year}{2012}).

\bibitem[{\citenamefont{Mathis et~al.}(1977)\citenamefont{Mathis, Rumpl, and Nordsieck}}]{Mathis77}
\bibinfo{author}{\bibfnamefont{J.~S.} \bibnamefont{Mathis}}, \bibinfo{author}{\bibfnamefont{W.}~\bibnamefont{Rumpl}}, \bibnamefont{and} \bibinfo{author}{\bibfnamefont{K.~H.} \bibnamefont{Nordsieck}}, \bibinfo{journal}{The Astrophysical Journal} \textbf{\bibinfo{volume}{217}}, \bibinfo{pages}{425} (\bibinfo{year}{1977}).

\bibitem[{\citenamefont{Ormel et~al.}(2007)\citenamefont{Ormel, Spaans, and Tielens}}]{Ormel07}
\bibinfo{author}{\bibfnamefont{C.~W.} \bibnamefont{Ormel}}, \bibinfo{author}{\bibfnamefont{M.}~\bibnamefont{Spaans}}, \bibnamefont{and} \bibinfo{author}{\bibfnamefont{A.~G. G.~M.} \bibnamefont{Tielens}}, \bibinfo{journal}{Astronomy \& Astrophysics} \textbf{\bibinfo{volume}{461}}, \bibinfo{pages}{215} (\bibinfo{year}{2007}).

\bibitem[{\citenamefont{Cuzzi et~al.}(1993)\citenamefont{Cuzzi, Dobrovolskis, and Champney}}]{Cuzzi93}
\bibinfo{author}{\bibfnamefont{J.~N.} \bibnamefont{Cuzzi}}, \bibinfo{author}{\bibfnamefont{A.~R.} \bibnamefont{Dobrovolskis}}, \bibnamefont{and} \bibinfo{author}{\bibfnamefont{J.~M.} \bibnamefont{Champney}}, \bibinfo{journal}{Icarus} \textbf{\bibinfo{volume}{106}}, \bibinfo{pages}{102} (\bibinfo{year}{1993}), ISSN \bibinfo{issn}{0019-1035}.

\bibitem[{\citenamefont{Cuzzi}(2004)}]{Cuzzi2004}
\bibinfo{author}{\bibfnamefont{J.~N.} \bibnamefont{Cuzzi}}, \bibinfo{journal}{Icarus} \textbf{\bibinfo{volume}{168}}, \bibinfo{pages}{484} (\bibinfo{year}{2004}), ISSN \bibinfo{issn}{0019-1035}.

\end{thebibliography}

\newpage 

\clearpage

\end{document}